\documentclass[preprint,superscriptaddress,floatfix,prl]{revtex4}
\usepackage[version=3]{mhchem} 
\usepackage{units}
\usepackage{graphicx,epsfig}
\newcommand{\HPc}{H$_2$Pc }

\newcommand{\EF}{$E_F$ }

\begin{document}
\title{Magnetoresistance through a single molecule}
\author{Stefan Schmaus}
\affiliation{Physikalisches Institut, Karlsruhe Institute of Technology (KIT),
 76128 Karlsruhe, Germany}
\affiliation{DFG-Center for Functional Nanostructures, Karlsruhe Institute of Technology (KIT),
 76128 Karlsruhe, Germany}

\author{Alexei Bagrets}
\affiliation{DFG-Center for Functional Nanostructures, Karlsruhe Institute of Technology (KIT),
 76128 Karlsruhe, Germany}
\affiliation{Institute of Nanotechnology, Karlsruhe Institute of Technology (KIT),  76128 Karlsruhe, Germany}

\author{Yasmine Nahas}
\affiliation{Physikalisches Institut, Karlsruhe Institute of Technology (KIT), 76128 Karlsruhe, Germany}
\affiliation{DFG-Center for Functional Nanostructures, Karlsruhe Institute of Technology (KIT),
 76128 Karlsruhe, Germany}

\author{Toyo K. Yamada}
\affiliation{Physikalisches Institut, Karlsruhe Institute of Technology (KIT), 76128 Karlsruhe, Germany}
\affiliation{Graduate School of Advanced Integration Science, Chiba University, Chiba 263-8522, Japan}

\author{Annika Bork}
\affiliation{Physikalisches Institut, Karlsruhe Institute of Technology (KIT), 76128 Karlsruhe, Germany}

\author{Martin Bowen}
\affiliation {Institut de Physique et Chimie des Mat\'eriaux de Strasbourg, UMR 7504 UdS-CNRS, 
67034 Strasbourg Cedex 2, France}

\author{Eric Beaurepaire}
\affiliation {Institut de Physique et Chimie des Mat\'eriaux de Strasbourg, UMR 7504 UdS-CNRS, 
67034 Strasbourg Cedex 2, France}

\author{Ferdinand Evers}
\affiliation{Institute of Nanotechnology, Karlsruhe Institute of Technology (KIT), 
76128 Karlsruhe, Germany}
\affiliation{Institut f\"ur Theorie der Kondensierten Materie, Karlsruhe Institute of Technology (KIT), 
D-76128 Karlsruhe, Germany}

\author{Wulf Wulfhekel}
\affiliation{Physikalisches Institut, Karlsruhe Institute of Technology (KIT), 76128 Karlsruhe, Germany}
\affiliation{DFG-Center for Functional Nanostructures, Karlsruhe Institute of Technology (KIT),
76128 Karlsruhe, Germany}


\maketitle

{\bf
The use of single molecules to design electronic devices
is an extremely challenging and fundamentally different approach to further downsizing
electronic circuits. Two-terminal molecular devices such as diodes were first 
predicted \cite{Aviram1974} and, more recently, measured experimentally \cite{Elbing2005}. 
The addition of a gate then enabled the study of molecular 
transistors \cite{McEuen,Delft,Natelson05}. In general terms, in order to increase data 
processing capabilities, one may not only consider the electron's charge but also its 
spin \cite{Wolf2001, Zutic2004}. This concept has been pioneered in giant 
magnetoresistance (GMR) junctions that consist of thin metallic
films~\cite{Baibich1988, Binasch1989}. Spin transport across molecules, 
i.\,e.\ \emph{Molecular Spintronics} remains, however, a challenging endeavor.
As an important first step in this field, we have performed an experimental and 
theoretical study on spin transport across a molecular GMR junction consisting of 
two ferromagnetic electrodes bridged by a \emph{single} hydrogen
phthalocyanine (H$_2$Pc) molecule. We observe that even though \HPc in itself is nonmagnetic, 
incorporating it into a molecular junction can enhance the magnetoresistance by one order 
of magnitude to 52\%. }

Scanning Tunneling Microscopy (STM) has proven to be a powerful and versatile tool for
studying electron transport properties of single molecules, be it by continually 
addressing individual molecules on a one by one level \cite{Joachim1995, Haiss2006, 
Neel2007, Takacs2008}, or in an automated break junction 
protocol \cite{Venkataraman06,Wandlowski2007}.
In this work, we introduce 
spin-polarized STM (Sp-STM) to measure
the magnetoresistance of single molecules, employing spin-polarized electrodes.
Experiments were carried out in a home-built STM working in ultra high vacuum
at 4 K \cite{Balashov2006} on individual \HPc molecules (C$_{32}$H$_{18}$N$_8$) 
sandwiched between Co-coated W tips and ferromagnetic Co nano-islands on Cu(111) single crystals.

\begin{figure}[t]
\centering
\includegraphics[width=0.85\linewidth]{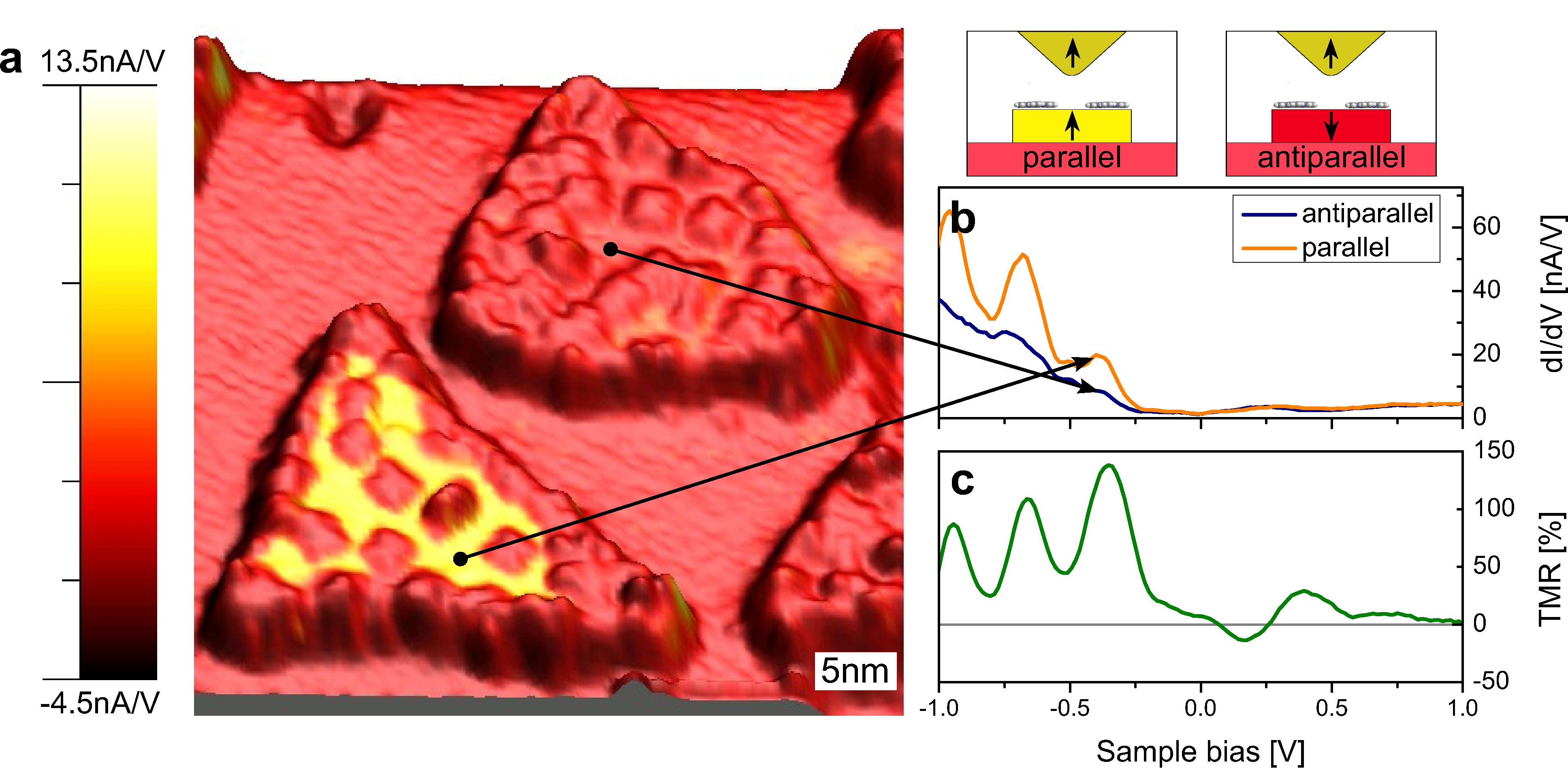}
\caption{a) Topographic image of \HPc molecules adsorbed onto two Co islands on the
Cu(111) surface. The colour code displays the measured d$I$/d$V$ map at
-310\,mV. One can distinguish between the two islands species with a 
magnetization parallel (P; in yellow) and antiparallel (AP; in red) with respect to the
tip magnetization. b) d$I$/d$V$ spectra taken on two islands of P and AP type, 
which clearly reveal spin-polarized states below the Fermi edge. c) Optimistic TMR ratio calculated from
the d$I$/d$V$ spectra. The highest value is measured at
around -350\,meV, and is used to distinguish between the two islands species.
}
\label{3dPlot}
\end{figure}

In Fig.\ \ref{3dPlot} we present the STM topography of the
sample after depositing \HPc molecules, that identify themselves by their four aromatic 
isoindole (BzPy) side groups \cite{Lippel1989,
Takacs2008}. We note that the Co nano-islands exhibit a spontaneous out-of-plane magnetization
due to a strong surface anisotropy \cite{Pietzsch2004}.
By using Co-coated tips (10 monolayers) with an out-of-plane magnetization, we can 
use the Sp-STM technique's sensitivity to the spin-polarized density of 
states \cite{Tersoff1985} to determine the relative orientation (parallel:P or 
antiparallel:AP) of the magnetization of individual islands relative to that of the tip.
Fig.\ \ref{3dPlot}b shows the differential conductance (d$I$/d$V$) curves measured 
atop two islands with P and AP alignments with respect to the tip. Particularly large 
differences in the spectra are found at
-350\,meV, which corresponds to the surface state of Co \cite{Diekhoener2003}.
This difference allows us to detect the local
magnetization direction of Co islands by recording maps of the local
differential conductance at this bias voltage (see Fig.\ \ref{3dPlot}a).
The differential tunneling magnetoresistance (differential TMR), defined as the difference 
in conductance divided by the smaller conductance of the
tunneling junction formed by the tip and sample, is strongly energy-dependent
as depicted in Fig.\ \ref{3dPlot}c. We note here that the differential TMR for 
this Co/vacuum/Co junction is only $\sim$\,5\,\% at low bias voltage.

\begin{figure}[t]
\centering
\includegraphics[width=0.65\linewidth]{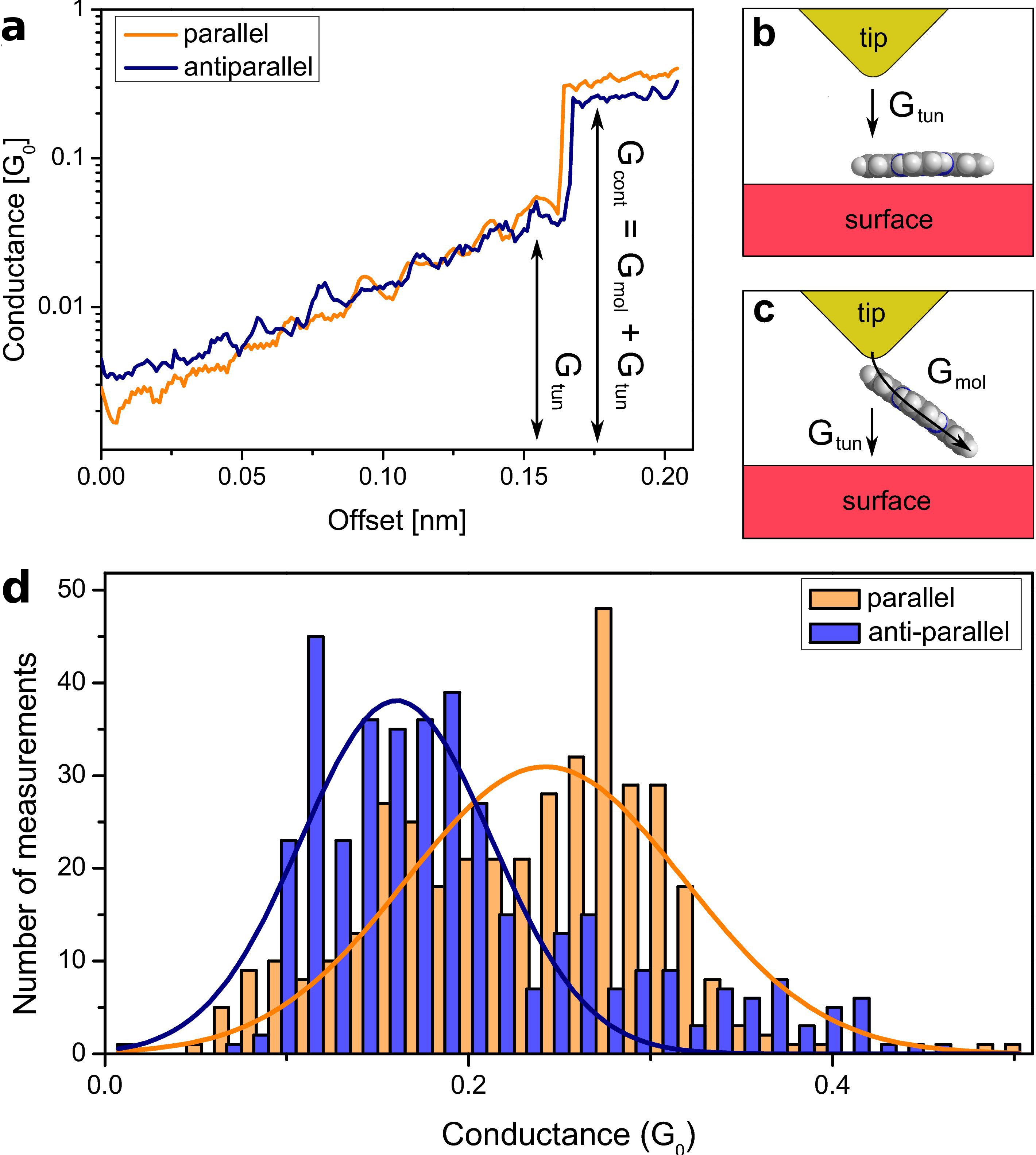}
\caption{a) A typical set of conductance-distance curves measured atop a \HPc 
molecule adsorbed onto P and AP magnetized
islands with a constant tunneling
voltage of 10\,mV ($G_0=\frac{2e^2}{h}$). As the tip approaches the molecule, 
the tunnel barrier width decreases, so that the conductance increases exponentially (see panel b)). Below a
certain tip-to-surface separation --- typically 3-4\,\AA{} --- the conductance abruptly 
increases as the molecule jumps into contact (see panel c), and then varies only slightly 
upon further reducing the distance.
Transport across the contacted molecule reflects both direct tunneling between the 
tip and the surface and conduction across the molecule (see panel c)). d) Histogram of the corrected
molecular conductances (390\,times parallel / 384\,times antiparallel). A Gaussian fit 
is used to determine the statistical conductance in the P and AP configurations, and 
thus the GMR ratio.}
\label{dc}
\end{figure}

To contact single molecules, we positioned the STM-tip above
the aromatic side groups of a \HPc molecule, opened the feedback
loop and decreased the tip-to-molecule distance ($\approx$~1\,\AA{}/s)
while measuring the tunneling current. We thus obtain conductance-distance curves
\cite{Joachim1995, 
Haiss2006, Neel2007}. Since the observed conductances of
phthalocyanine molecules are rather high \cite{Takacs2008}, a small
voltage (10\,mV) has to be used to avoid the thermal disintegration of the
molecules. We observe an exponential increase of the conductance that reflects the decreasing tunneling barrier width
(see Fig.\ \ref{dc}a). Below a certain tip-to-surface separation --- 
typically 3-4\,\AA{} --- the conductance abruptly increases, which indicates a sudden change
of the junction geometry.
Similar jumps have been seen, \textit{e.\,g.}, for
alkanedithiol molecular wires \cite{Haiss2006}. We have recently shown that,
for phthalocyanine molecules, this corresponds to a lifting of the flat \HPc molecule's
aromatic group as it contacts the tip \cite{Takacs2008}.

The conductance after the molecular jump-to-contact depends only slightly on the distance and encompasses both the
transport across the molecule $G_{\mathrm{mol}}$ and direct tunneling between the tip and the sample
$G_{\mathrm{tun}}$ \cite{Haiss2006} (see Fig.\ 2c). This conductance $G$ in molecular contact mode
depends markedly on whether the underlying island is of P or AP type. In the particular measurement
of Fig.\ \ref{dc}a, we find
$G_\mathrm{P}= 0.30\, G_0$ and $G_{\mathrm{AP}}= 0.17\, G_0$) measured at 10\,mV bias, 
where $G_0 = \tfrac{2e^2}{h}$ is the quantum of conductance.
We may eliminate a sizeable proportion of the spin-dependent direct (tip-to-island)
tunneling contribution to the conductance $G$ by subtracting the measured conductance $G_{\mathrm{tun}}$ before the
jump from that after $G_{\mathrm{cont}}$, \textit{i.\,e.}\ $G_{\mathrm{mol}} \approx G_{\mathrm{cont}}-G_{\mathrm{tun}}$.

To ensure identical tip conditions when quantitatively comparing P and AP conductances, measurements on two Co
islands of P and AP type were performed in the same
scan. This approach further eliminates magnetostriction of the two ferromagnetic electrodes. 
Each such measurement was in turn repeated several hundred times and
the distribution of the P and AP conductances is depicted in
Fig.\ \ref{dc}d. The width of the conductance distribution
-- which underscores the noise of the conductance measurement and a variation in contact
geometries -- is relatively small when compared to
previous work \cite{Haiss2006}; the histogram of the measurements on nearly 800 P and AP 
junctions clearly reveals a difference in conductance between the P and AP junctions.
Gaussian fits were used to determine the GMR from the measurement statistics.
We find
$G_{\mathrm{P}}= (0.242 \pm 0.007)\,G_0$ and
$G_{\mathrm{AP}}=(0.160 \pm 0.005)\,G_0$. These values result in an
optimistic GMR ratio
of $$\text{GMR}=\frac{G_{P}-G_{AP}}{G_{AP}}=(51\pm9)\%\text{.}$$
Surprisingly, the GMR ratio obtained at V\,=\,10\,mV is one order of magnitude larger 
than the differential TMR found for direct tunneling between the tip and the Co surface.

To understand what causes this large value of GMR, we have
performed transport calculations based on density functional theory (DFT)
employing the nonequilibrium Green's function (NEGF) formalism
and the TURBOMOLE package \cite{arnold2007}; details may be found in the Supplementary Information.

\begin{figure}[t]
\includegraphics[width=0.5\linewidth]{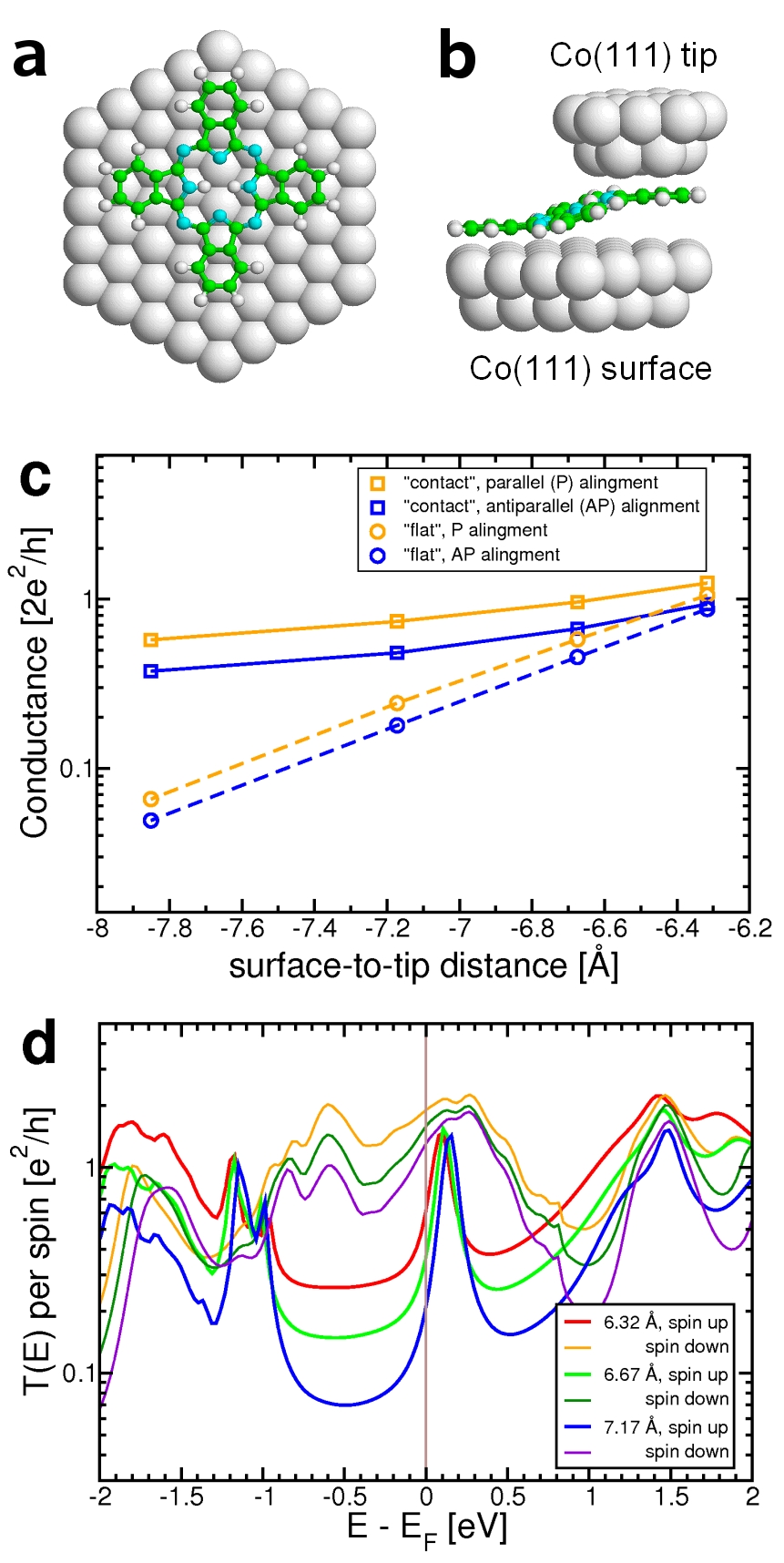}
\caption{a),b) Contact geometry used in the transport calculation for a \HPc molecule a) 
adsorbed on the Co-island and b) in simultaneous contact with the tip and the Co surface 
through a lifting of the aromatic group.
Cobalt sites are in grey; hydrogen in white; carbon in green; nitrogen in cyan.
c) Conductance of \HPc sandwiched between two P- or AP-aligned Co(111) surfaces.
The upper(lower) pair of traces corresponds to the "contact"("flat") junction geometry.
d) The transmission probability $T(E)$ of an electron with energy
$E$ through a molecular junction of P type for the majority (lower three traces) 
and minority (upper three traces) spin channels.}
\label{f3}
\end{figure}

To find the atomic structure of the molecular junction,
we first performed a geometry optimization for \HPc
on the Co(111) surface, represented by a cluster with 65 atoms.
Our analysis suggests that \HPc adsorbs preferentially in the bridge position onto 
Co(111) (see Fig.\ \ref{f3}a), owing to a binding energy $-8.17$~eV that is larger than 
that found in either the hollow site position ($-8.06$~eV) or the atop site position 
($-7.45$~eV), consistent with earlier findings \cite{Iacovita2009, Heinrich2010}.

We have calculated spin-polarized transport in the linear response at low bias voltage for the two junction
geometries schematized in Fig.\ \ref{f3}a,b) before ("flat") and after ("contact") the jump-to-contact.
To establish
the latter configuration, a free \HPc molecule has been bent along the
low energy vibrational eigenmode with frequency $\omega = 6.4$~meV \cite{Takacs2008}.
Within the NEGF calculations, the magnetization direction (P or AP relative to the tip)
of each Co-cluster is a control parameter. 
We have ascertained that, in our study, the electronic
structure of the Co surface is properly reproduced with, in particular, an 
exchange splitting of the $d$-states of 1.8~eV (see Supplementary Information). This leads to
a magnetic moment $m$(Co)$\ \approx 1.65\,\mu_B$ per surface Co atom,
in agreement with previously reported calculations~\cite{ABagrets2007}.

On a qualitative level, our transport calculations (see Fig.\ \ref{f3}c) reproduce very well our
experimental findings (see Fig.\ \ref{dc}a). We confirm the exponentially
increasing conductance in the tunneling regime, $G_{\rm tun}(d) \sim e^{-\beta d}$,
for which the distance $d$ between the two electrodes is still large. The slope is independent of the
relative alignment of electrode magnetizations and the computational value
$\beta^\mathrm{theo} = 1.87$~\AA$^{-1}$ (\textit{i.\,e.}\ the work function
$W^\mathrm{theo} = 3.24$~eV) is in agreement with experiment ($\beta=1.9\pm0.3$~\AA; $W = 3.2$~eV).

Once contact between the tip and the molecule has been established,
$G_{\rm cont}(d)$ varies much more weakly with the contact
distance $d$ just as observed in the experiment. It is, however, still sensitive to the relative
orientation of the magnetization of the electrodes. We find that
$G_\mathrm{AP}(d)$ is always much lower than
$G_\mathrm{P}(d)$. For a quantitative comparison with
experiment, we consider the GMR$^{\rm theo}$ ratio at the distance $d$
for which the ratio $r=G_{\rm tun}/G_{\rm cont}$
matches the value $r\approx 4$ found experimentally.
We thus find that GMR$_{\rm theo}\approx 65$~\% and is only
weakly dependent on $d$.

We now discuss the conduction mechanism from the substrate onto the molecule and 
across the molecular contact established between the tip and the aromatic group, and 
its impact on the large GMR measured. Pc molecules are characterized by an energetically isolated highest occupied
molecular orbital (HOMO) and a
nearly doubly degenerate lowest unoccupied
molecular orbital (LUMO) \cite{Rosa2001}.
We note that the HOMO$^*$ levels corresponding to the aromatic group hybridize only very weakly,
with almost no amplitude on the bridging
nitrogen (N$_{\rm b}$)$_4$.
By contrast, the LUMO states are located on two out of the four aromatic groups, 
with a strong hybridization to all N atoms forming the inner macro-cycle 
(see Supplementary Information). Since the N bond to Co includes states at the Fermi energy 
\EF \cite{Takacs2008}, transport should occur via the
quasi-degenerate LUMO level. We confirm this fact by examining in Fig.\ \ref{f3}d) the transmission probability
per spin direction $T_{\uparrow,\downarrow}(E)$
at the Fermi energy
across a junction of P type.
We find that $G_{\mathrm{P}}$ near \EF is indeed weighted by a peak centered slightly above
\EF that underscores transmission through the LUMO level. As we discuss in the 
Supplementary Information, the larger density of Co minority states at \EF results, through 
the N-Co bond, in a larger efficiency of LUMO hybridization, and thus of LUMO broadening, 
in the spin $\downarrow$ channel.

This difference in LUMO broadening for the two spin channels has a direct impact on the 
GMR measured across the molecular junction. Indeed, the conductance across a single level, 
here the LUMO, generically takes on the Breit-Wigner form \cite{Huisman2009}
$G\approx \Gamma^\text{substrate}\Gamma^\text{tip}/( (E_\text{LUMO}-E_\text{F})^2 +
(\Gamma^\text{substrate}+\Gamma^\text{tip})^2/4)$, which considers
the energy separation between the LUMO and \EF, as well as the LUMO
broadenings (inverse lifetimes) $\Gamma^\text{substrate}$ and $\Gamma^\text{tip}$ due to 
hybridization to the substrate and tip, respectively. Each is in turn split into 
$\Gamma^{\rm min(maj)}$ depending on the spin channel considered.
Because transport is off-resonant, i.e.
$|E_\text{LUMO}-E_\text{F}|\gg \Gamma^\text{min,maj}$,
we have
$G_{\mathrm{P}}\approx (\Gamma^{\rm min}\Gamma^{\rm min} + \Gamma^{\rm
maj}\Gamma^{\rm maj})/(E_{\rm LUMO}-E_{\rm F})^2$
while
$G_{\mathrm{AP}}\approx 2 \Gamma^{\rm min}\Gamma^{\rm
maj}/(E_{\rm LUMO}-E_{\rm F})^2$.
Introducing the ratio $\varrho=\Gamma^{\rm
maj}/\Gamma^{\rm min}$ we thus find
$$
{\rm GMR} \approx \frac{(\Gamma^{\rm min}-\Gamma^{\rm maj})^2}{2\Gamma^{\rm
  min}\Gamma^{\rm maj}} = \frac{(1-\varrho)^2}{2\varrho}.
$$
This simple formula implies two important rules of thumb for spin-polarized transport off-resonance across a molecule.
First, the GMR is insensitive to the precise location of the resonance energy.
Second, it is mainly indicative of the {\rm ratio} $\varrho$ of minority and
majority molecular orbital broadenings due to hybridization. Relative to the small 
differential TMR ratio found between the Co substrate and tip, the larger GMR ratio 
measured experimentally can be explained theoretically by the above ratio $\varrho$. 
The imbalance within the two spin channels of the hybridization-induced level broadening 
to the molecular orbital responsible for transport promotes a high GMR across the molecule.

In conclusion, we have experimentally demonstrated a GMR of over 50\,\% in single
molecule junctions that is much larger than the value of differential TMR found without
the presence of molecules in the junction. Such a high value is caused by a strong 
hybridization of the molecular LUMO, responsible for transport, with minority states 
of the two metallic electrodes. Such a selective hybridization thus leads to  a spin 
filtering effect that could be generic to all molecular junctions.

We express our gratitude to O.\ Hampe, J.\ Kortus, K.\ Fink, S.\ Boukari, Xi Chen, 
M.\ Alouani, R.\ Mattana, J.\ van Ruitenbeek and P.\ Seneor for useful communications 
and acknowledge support by the DFG (WU 349/3-1 and SPP1243), the Alexander von 
Humboldt foundation and the ANR (ANR-06-NANO-033-01).
\bigskip

{\bf Methods}

The Cu(111) crystal was cleaned thanks to several cycles
of Ar$^{+}$ sputtering and annealing.
The molecules were evaporated in situ from a Knutsen cell heated to
$\approx$\,500\,K. During the deposition process we keep the
sample at 270\,K to reduce thermal diffusion of the
deposited molecules.
d$I$/d$V$ curves were
measured on the bare islands with the lock-in technique.

DFT-based transport calculations
were carried out with a homemade code building upon the
NEGF formalism and the TURBOMOLE package \cite{arnold2007}.
Our implementation enables us to perform transport simulations
with free boundary conditions, which for the present case have been
extended to account for the spin-polarized electronic structure
of the magnetic electrodes (for further details, see Supplementary Information).
The gradient corrected approximation (GGA)
DFT-energy has been amended by empirical corrections \cite{Grimme}
to account for dispersive van der Waals
interactions between the molecule and the surface.

\bibliographystyle{nature}
\bibliography{nano}

\clearpage

\clearpage

\clearpage

\clearpage

\renewcommand{\figurename}{Suppl.\ Fig.} 
\setcounter{figure}{0}

\subsection*{{\large SUPPLEMENTARY INFORMATION: }}
\subsection*{{\large Magnetoresistance through a single molecule}}

Stefan Schmaus, Alexei Bagrets, Yasmine Nahas, Toyo K. Yamada 

Annika Bork, Martin Bowen, Eric Beaurepaire, Ferdinand Evers, and Wulf Wulfhekel

\subsection*{1.\ Geometry optimization}
We used the quantum chemistry package\ {\small TURBOMOLE}~[Ref.1] 
to find the atomic structure of molecular junctions. We have 
performed standard geometry optimization for \HPc on Cu(111) surface, 
represented by a cluster with 65 atoms. (The use of Co(111)-clusters for structure 
optimization is impaired by the very large number of energetically almost 
degenerate spin multiplets.) The gradient corrected approximation (GGA~[Ref.2]) 
DFT-energy has been amended by empirical corrections~[Ref.3] 
to account for dispersive van der Waals 
interactions between the molecule and the surface. 
Our analysis suggests, that \HPc prefers the "bridge" position 
(see the paper, Figure 3a) with binding energy $-8.17$~eV, against the 
"hollow" site position (binding energy $-8.06$~eV) and 
"atop" site position (binding energy $-7.45$~eV), 
which is consistent with earlier findings~[Refs.4,5].

\subsection*{2.\ Electronic structure and transport calculations 
through H$_2$Pc}

\begin{figure*}
\includegraphics[width=0.85\linewidth]{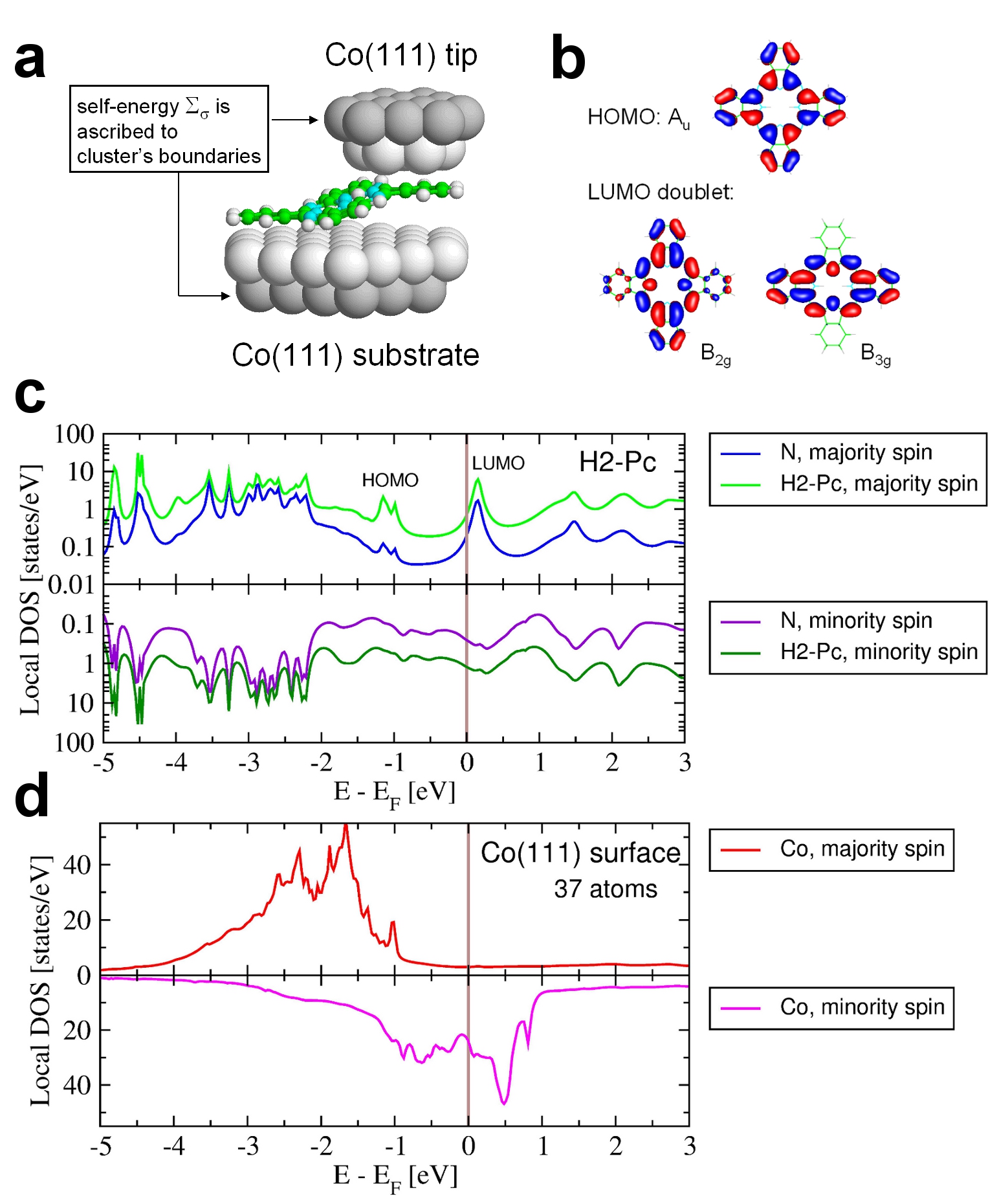}
\caption{(a) Atomic cluster, representing a bottle-neck of the H$_2$Pc molecular junction, 
used for the electronic structure and transport calculations; outermost
boundaries of the Co clusters (dark gray) are subject to the absorbing boundary
conditions modeled by the self-energy $\Sigma_\sigma$. (b) Frontier molecular
orbitals of H$_2$Pc (D$_{2h}$ symmetry): an A$_u$ HOMO and a quasi-degenerate LUMO doublet,  
B$_{2g}$ and B$_{3g}$. (c) and (d): spin-polarized local density of states
at H$_2$Pc and Co(111) surface, respectively. }
\label{supp_fig1}
\end{figure*}

Density functional theory (DFT) based electronic structure and
transport calculations through \HPc molecular junctions have 
been performed within the non-equilibrium Green's function (NEGF)
approach as implemented in a homemade simulation code~[Ref.6] 
interfaced to the quantum chemistry package\ {\small TURBOMOLE}~[Ref.1].
The atomic configuration of the "extended molecule" used to 
simulate a bottle-neck of the molecular junction is
shown in Suppl.~Fig.~\ref{supp_fig1}a ("contact" regime): \HPc is bound to the two Co(111) clusters
with 51 and 19 atoms, representing the Co surface and the Co STM-tip, respectively.
To establish this configuration a free \HPc molecule has been bent along the 
low energy vibrational eigenmode with frequency $\omega = 6.4$~meV.
A similar atomic configuration, with \HPc bound to the surface only, has been
used for transport simulations in the tunneling regime (see the paper, Figure 3b). 
The generalized gradient approximation (GGA, BP86 functional~[Ref.2]) 
and a contracted Gaussian-type split-valence basis set 
with polarization functions (SVP)~[Ref.7] 
have been employed for calculations. 

First, a closed-shell (nonmagnetic) solution for the "extended molecule"
comprising 2156 electrons has been found. 
To account for infinite reservoirs with spin-polarized electrons,   
the spin-dependent ($\sigma = \pm 1/2$) local self-energies, 
$
 \Sigma^{\mathrm{surface/tip}}_{\sigma}(\mathbf{x,x'}) = 
 \left[\lambda + \sigma \Delta_\mathrm{ex} - i\gamma \right] \delta(\mathbf{x} - \mathbf{x'}),
$
have been ascribed to the outermost boundaries of the simulation cluster 
(dark gray atoms at Suppl.~Fig.~\ref{supp_fig1}a). 
Here, a parameter $\Delta_\mathrm{ex} = 1.6$~eV accounts for 
exchange splitting of the bulk Co $d$-states~[Ref.8]. 
A freedom to choose a sign of $\Delta_\mathrm{ex}$
independently for the "surface" and "STM-tip" clusters allows for two solutions: 
with parallel and antiparallel alignment of electrodes' magnetizations.  
The non-equilibrium Green's
function formalism is employed to evaluate the charge- and spin-density matrices 
in the presence of open boundaries
insuring a charge neutrality within the "extended molecule".
The density matrices are given back to  
{\small TURBOMOLE} to find a modified set of Kohn-Sham orbitals, 
with a cycle to be repeated unless the self-consistent solution is reached.
For the given value of the level broadening, $\gamma$ ($\ = 1.36$~eV in present 
calculations), the contribution $\lambda$ to the real piece of the self-energy has
been defined by imposing the condition of spurious charge accumulation to be 
absent at the cluster's boundaries. Further details of the computational approach will
be published elsewhere~[Ref.9].

As an example, details of electronic structure for the
molecular junction in the case of parallel alignment of magnetizations 
are presented in Suppl.~Fig.~\ref{supp_fig1}c,d. The electronic states at the Co-surface are 
spin-polarized with exchange splitting\ $\approx 1.8$~eV (Suppl.~Fig.~\ref{supp_fig1}d), 
giving rise to an average magnetic moment\ \mbox{$\approx 1.65 \mu_B$}\ per a surface Co atom. 
Inspecting the local density of states (DOS) at H$_2$Pc, Suppl.~Fig.~\ref{supp_fig1}c, we 
observe the majority spin resonance above the Fermi level ($E_F$) to be identified 
with the quasi-degenerate LUMO, which has a significant
weight on the nitrogen atoms (cf.\ Suppl.~Fig.~\ref{supp_fig1}b). 
In contrast, no density is seen on nitrogens for the majority spin 
HOMO resonance positioned $\sim 1.2$~eV below $E_F$ 
that reflects the stucture of the HOMO (Suppl.~Fig.~\ref{supp_fig1}b).
Furthermore, the minority spin H$_2$Pc LUMO level evolves into a broad
peak due to a hybridization with the minority spin Co states, which 
density near $E_F$ significantly exceeds the ones for majority spin electrons. 
The asymmetry in the LUMO level broadenings, 
$\Gamma^\mathrm{min} > \Gamma^\mathrm{maj}$,
gives rise to a magnetoresistance effect 
(see text of the paper for further details). 

\clearpage

\clearpage

\subsection*{{\large References (Supplementary Information)}}

{\small
\begin{description}
\itemsep 0pt
\item{[Ref.1]}
TURBOMOLE V5.10 by R.\ Ahlrichs {\it et al.}\ (www.turbomole.com)

\item{[Ref.2]}
(a) Becke A.\ D.\ {\it Phys.\ Rev.\ A} {\bf 1988}, {\it 38}, 3098;
(b) Perdew J.\ P.\ {\it Phys.\ Rev.\ B} {\bf 1986}, {\it 33}, 8822.

\item{[Ref.3]}
Grimme, S.\ {\it J.\ Comp.\ Chem.}\ {\bf 2006}, {\it 27}, 1787--1799.

\item{[Ref.4]}
Iacovita, C.; Rastei, M.; Heinrich, B. W.; Brumme, T.; Kortus, J.; Limot, L.\ \&
Bucher, J.\ {\it Phys.\ Rev.\ Lett.}\ {\bf 2009}, {\it 101}, 116602.

\item{[Ref.5]}
Heinrich, B.\ W.; Iacovita, C.; Brumme, T.; Choi, D.-J.; Limot, L.; Rastei, M.\ V.;
Kortus, J.; Hofer, W.\ A.\ \& Bucher, J.-P.\
{\it Selective bonding and apparent symmetry of single
Cobalt-Phtahalocyanine molecules on a copper (111) surface}. Preprint, {\bf 2010}.

\item{[Ref.6]}
Arnold, A.; Evers, F.\ \& Weigend, F.\
{\it J.\ Chem.\ Phys.}\ {\bf 2007}, {\it 126}, 174101.

\item{[Ref.7]}
Sch{\"a}fer, A.; Horn H.\ \& Ahlrichs, R.\
{\it J.\ Chem.\ Phys.}\ {\bf 1992}, {\it 97}, 2571.

\item{[Ref.8]}
Moruzzi, V.L.;\ Janak, J.\ F.\ \& Williams, A.\ R.\ {\it Calculated
Electronic Properties of Metals } (Pergamon Press, New York, 1978).

\item{[Ref.9]}
Bagrets A.\ {\bf 2009}, unpublished.

\end{description}
}

\end{document}